\documentclass{article}

\usepackage{booktabs}
\usepackage{graphicx}
\usepackage{url}
\usepackage{amsmath,amssymb}
\usepackage{caption}
\usepackage{subcaption}
\usepackage{hyperref}

\usepackage[top=1in, bottom=1.25in, left=1.25in, right=1.25in]{geometry}

\title{AdaptiFont: Increasing Individuals' Reading Speed with a Generative Font Model and Bayesian Optimization}
\author{Florian Kadner$^{\text{a},\text{b}}$ \qquad Yannik Keller$^{\text{a},\text{b}}$ \qquad Constantin A. Rothkopf$^{\text{a},\text{b}}$}
\date{$^{\text{a}}$Centre for Cognitive Science, TU Darmstadt, Darmstadt, Germany \\ \vspace{0.1cm}
$^{\text{b}}$Institute of Psychology, TU Darmstadt, Darmstadt, Germany}
\begin{document}
\maketitle
\begin{abstract}
Digital text has become one of the primary ways of exchanging knowledge, but text needs to be rendered to a screen to be read. We present AdaptiFont, a human-in-the-loop system that is aimed at interactively increasing readability of text displayed on a monitor. To this end, we first learn a generative font space with non-negative matrix factorization from a set of classic fonts. In this space we generate new true-type-fonts through active learning, render texts with the new font, and measure individual users’ reading speed. Bayesian optimization sequentially generates new fonts on the fly to progressively increase individuals’ reading speed. The results of a user study show that this adaptive font generation system finds regions in the font space corresponding to high reading speeds, that these fonts significantly increase participants’ reading speed, and that the found fonts are significantly different across individual readers.
\end{abstract}

\begin{figure}[h]
  \includegraphics[width=\textwidth]{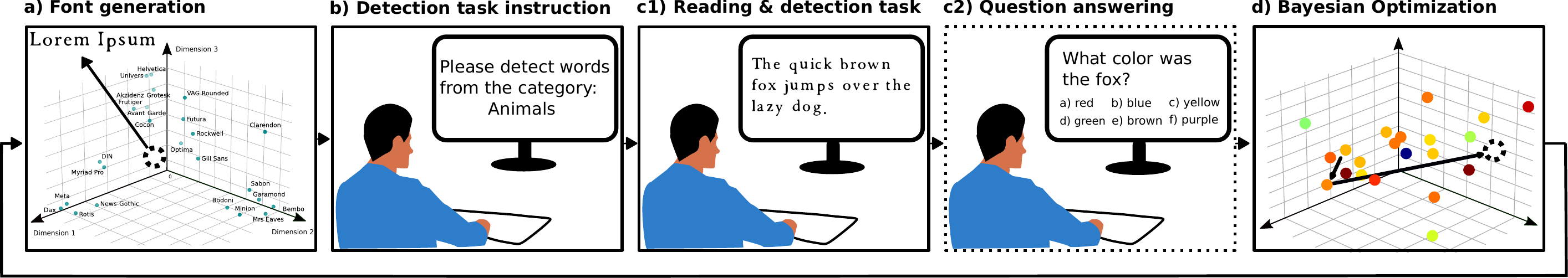}
  \caption{Schematic of the closed-loop algorithm for generating and optimizing fonts to increase individuals' reading speed.}
  \label{fig:teaser}
\end{figure}
\section{Introduction}
Language is arguably the most pervasive medium for exchanging knowledge between humans \cite{miller1991science}. But spoken language or abstract text need to  be made visible in order to be read, e.g. in print or on screen. 
Traditionally, the transformation of text to the actually visible letters follows a complex process involving artistic and design considerations in the creation of typefaces, the selection of specific fonts, and many typographic decisions regarding the spatial placement and arrangement of letters and words \cite{carter2011typographic,zapf1991classical}. 

A fundamental question concerning the rendering of text is whether the way text is displayed affects how it is being read, processed, and understood. 
Scientifically, this question lies at the intersection of perceptual science and cognitive science.
Numerous empirical investigations have measured how properties of written text influence its perception in reading both on paper \cite{legge1985psychophysics,Bigelow2019,landolt2019visual} and on electronic devices \cite{dyson2004physical,bernard2002comparison,bruijn1992influence}.
Nevertheless, overall results of these empirical investigations on the relationship between reading speed or reading comprehension and font parameters are mixed and in part contradictory. 

More recently, due to the pervasive use of electronic text that can be read on different devices under different viewing conditions, adaptive methods for font rendering and text display have been developed \cite{andre1994dynamic,sheedy2008cleartype,ms_opentype}. Moreover, fully parameterizable fonts intended to generate new fonts have been developed for multiple purposes \cite{hu1998synthesis,Arditi2004,devroye1995random,bragg2016reading}.
Some of these fonts are intended to help designing new fonts \cite{hu1998synthesis}, some have been developed to enhance readability for individuals with low vision \cite{Arditi2004}, and others have been developed to increase readability on small electronic devices \cite{bragg2016reading}.

Here, we close the loop by adaptively generating fonts and optimizing them progressively so as to increase individual's reading speed. 
The system is based on a generative font space, which is learnt in a data driven fashion using non-negative matrix factorization (NMF) of 25 classic fonts. We use this font space to generate new fonts and measure how a generated font affects reading speed on an individual user by user level. By utilizing Bayesian optimization, we sample new fonts from the generative font model to find those fonts that allow the reader to read texts faster. We demonstrate the feasibility of the approach and provide a user study showing that the found fonts indeed increase individual users' reading speed and that these fonts differ between individuals.

\section{Related work}
\subsection{Typography, Font Design, and Reading}
Traditionally, the development of a typeface is a complex creative process influenced by aesthetic, artistic, historic, economic, social, technological and other considerations \cite{carter2011typographic}. As such, designing of typefaces and typography underlie individual judgements, which may be highly variable between individuals and across time, see e.g. \cite{jarlehed2015typographic}. Nevertheless, readability has often been a central goal in the creation of typefaces, as expressed by Adrian Frutiger: "... optimum readability should always be foremost when developing a typeface" \cite{osterer2014adrian}. What constitutes readability in this context and how to measure legibility and readability, has been debated extensively \cite{dale1949concept,york2008legibility}.

\subsection{Typeface Features and Legibility Research}
The scientific and empirical investigation of the relationship between how rendered text looks and how it is processed lies at the intersection of cognitive science and perceptual science and is particularly relevant in psycholinguistics. The interested reader is refereed to reviews providing an overview of this broad body of work \cite{legge1985psychophysics,Bigelow2019,dyson2004physical,landolt2019visual}.
Of particular relevance for the present study are investigations addressing how font features such as size, boldface, serifs, or different typefaces affect text readability. In this context, readability has been operationalized in different ways
with subject's reading speed such as words per minute being the most common way. Note that within typography, the effect of a typeface's design and the glyph's shape is usually termed legibility and not readability.

Perceptual science has long investigated how features of typefaces and individual fonts affect letter perception e.g. with a series of experiments by Peterson and Tinker since the 1920ies \cite{paterson1931studies}. Since then, numerous features affecting readability of text have been investigated, including font size \cite{Legge2011}, serifs versus sans-serif typefaces \cite{Arditi2005}, lightness contrast \cite{blommaert1987letter}, color contrast \cite{mclean1965brightness} and many others.
Perceptual science has also investigated how the spacing between adjacent letters influence letter recognition. Crowding describes the phenomenon that the perception of letters depends on their spatial separation as a function of foveal eccentricity \cite{bouma1970interaction}. It has been shown that crowding can account for the parafoveal reading rate \cite{Pelli2007}. Similarly, the effect of the vertical spacing of words on readability has been investigated \cite{chung2004reading}. A number of studies have particularly focused on how readability of fonts changes with age \cite{darroch2005effect} and how to make fonts more readable for readers with medical conditions \cite{russell2007legibility}. 

Because of the importance of text in electronic communication, the field of HCI has investigated the effect of fonts on readability particularly when rendering text to monitors and other electronic devices. Such studies include investigations of the effect of font size and type on glance reading \cite{dobres2016utilising}, 
measurement of user preferences and reading speed of different online fonts an sizes \cite{boyarski1998study,bernard2002comparison}, 
comparisons of the effect of font size and line spacing on online readability \cite{rello2016make}, quantifying the interaction of screen size and text layout \cite{bruijn1992influence}, and investigating feature interaction such as color and typeface in on-screen displays \cite{garcia1996effect}.

Nevertheless, integrating all these results provides a mixed picture. Several studies concluded that increasing font size improves readability \cite{rello2016make} while others reported that this holds only up to a critical print size \cite{chung1998psychophysics}. While early studies reported improvements in readability with increased font weight \cite{luckiesh1940boldness}, other studies reported no significant effect \cite{bernard2013effect}. While the presence of serifs did not significantly affect the speed of reading \cite{Arditi2005}, font type did but only modestly \cite{mansfield1996psychophysics}.
Complicating the interpretation of some of these results is that most studies analyzed reading data across subjects, while some studies found more complex interactions of font features at the individual subject level
\cite{Korinth2020}. Similarly, it remains unclear how reading text in print relates to reading on screen \cite{kopper2016reading}. Finally, reading research has utilized many different reading tasks ranging from RSVP letter detection to comprehension of meaningful text \cite{salcedo1972broader}.

\subsection{Parametric, Adaptive, Generative, and Smart Fonts}
The introduction of digital font has generated new techniques and challenges for adaptively rendering and setting text particularly on electronic devices, both from the point of view of the development of such algorithms \cite{knuth1999digital} and from the point of view of the typographer \cite{zapf1991classical}. Parametric tools intended to help designers in developing new fonts have been introduced, e.g. \cite{shamir1998feature}.
Parametric techniques have been developed with the aim to increase readability, 
both in print \cite{andre1994dynamic} as in rendering on screen, e.g. ClearType adjusts sub-pixel addressing in digital text to increase readability on small displays \cite{sheedy2008cleartype}. 
Knuth's metafont goes a step further, as it can specify a font with one program per glyph, which in the case of the Computer Modern font family generates 72 fonts through variation of parameters. 
More recent developments include 
OpenType, which uses multiple glyph outline that are rendered to variable parametric fonts \cite{ms_opentype} as well as adaptive fonts, which have recently gained popularity in responsive design for the web \cite{matej2019better}.
The ubiquity of hypertext has similarly lead to adaptive grid based layouts \cite{jacobs2003adaptive}, which affects readability of text. Some of these systems are based on optimization algorithms that try to formulate and include font parameters \cite{hurst2009review}.

On the other hand, the development of computational and algorithmic art \cite{nake2009semiotic} and design \cite{fishwick2008aesthetic} have extended the possibilities of creation in many fields.
Parameterized generative font models allowing the parametric generation of new fonts have been created \cite{mcqueen1993infinifont,hu1998synthesis}. 
Other approaches aimed at describing characters as an assembly of structural elements resulting in a collection of parameterizable shape components \cite{hu2001parameterizable}. A prototype parametric font program called Font Tailor was specifically designed for users with visual impairments to customize a font to increase readability \cite{Arditi2004}. 
Dynamic fonts' \cite{andre1990dynamic} shapes are instantiated every time a glyph is rendered allowing for parameterized variability, e.g. in emulating handwriting \cite{devroye1995random}, and fluid typography \cite{brownie2007one} changes glyphs' shape on screen thus blurring the lines between typography and animation. Some dynamic fonts have been design specifically to increase readability on small portable electronic devices \cite{bragg2016reading}.

Our data driven approach to obtain a generative font space is most closely related to previous approaches of unsupervised learning of fonts \cite{campbell2014learning} and recent systems based on Generative Adversarial Networks (GAN), e.g. \cite{GAN1,GAN2, GAN3, GAN4, GAN5}. Differently from \cite{campbell2014learning}, which used a polyline representation of letters, we used a pixel based representation as the current study renders text only to a monitor up to a size of 40 pt. 

\subsection{Adaptive Design through Optimization in HCI}
Recent work has explored the use of optimization methods from machine learning to evolve and adapt designs with explicit and implicit measures of users' preferences, which in general can be seen as a field belonging to interactive machine learning \cite{fails2003interactive}. The difficulty lies in the availability or construction of an appropriate quantitative criterion that can be maximized and captures humans' explicit or implicit preferences. 
Note that this is different from the supervised learning of selecting fonts from many design examples \cite{zhao2018modeling}.
One such area is the automatic design of interface layouts. 
Some approaches optimize layouts with respect to hand coded rules that are aimed at incorporating design criteria \cite{purvis2003creating}.
Off-line systems collect a large data set of user preferences or abilities and then approximate the user through a function, e.g. in ability-based user interfaces \cite{gajos2008improving,gajos2010automatically}.
More recent work trained a neural network to predict users' task performance from a previously collected data \cite{duan2020optimizing}. 
Closed-loop adaptive systems, i.e. a system that parameterically changes to optimize some interaction criterion on-line while the user is actively interacting with it, are much rarer and have been used predominantly in the context of game design \cite{ZookFR14,raffert2012optimally,mahmud2014adapting}. 

\subsection{Bayesian Optimization in HCI}
One particularly attractive and powerful method for optimizing an unknown function is Bayesian optimization. It has a long tradition in many fields of research including optimal design of experiments \cite{shahriari2015taking} including closed-loop design of experiments \cite{lorenz2016automatic}. The power of Bayesian optimization lies in its ability to use statistical methods in modeling and efficiently optimizing ‘‘black-box’’ functions. Bayesian optimization has been used in several recent HCI systems both for open-loop optimization \cite{dudley2019crowdsourcing} as well as in closed-loop systems, e.g. in the context of computer game design \cite{khajah2016designing}. Current research also includes the application of Bayesian optimization in the field of computational design involving crowdsourcing \cite{Koyama2018}.

\section{Methods}

\subsection{Preliminary User Study}
In a preliminary user study we wanted to test whether reading speed was related to typefaces and font parameters at an individual subject's level.
Seven subjects participated and read 250 texts each, whereby their reading speed was measured. All subjects had normal or corrected to normal vision and were German native speakers. They were seated at a distance of approximately 50 cm in front of a 24 inch monitor, which had a resolution of 1080 pixels horizontally. Participants were instructed to read the texts attentively and as quickly as possible, but only once in total.

\begin{table}[!b]
\caption{Bayes Factors of the ANOVA relating font features and individuals' reading speed.}
\label{tab:preliminary}
\begin{minipage}{\textwidth}
\begin{center}
\resizebox{0.75\textwidth}{!}{\begin{tabular}{llllllll}
                        & 1     & 2     & 3    & 4    & 5     & 6     & 7     \\
\toprule
Font Size               & \textbf{4.34}  & \textbf{20.22} & 0.81 & \textbf{6.08} & 1.27  & \textbf{12.69} & \textbf{40.80} \\
Font Name               & \textbf{3.53}  & 0.46  & 1.15 & 0.54 & 0.16  & 0.55  & 0.01  \\
Bold                    & 1.20  & 0.69  & \textbf{6.09} & 0.14 & \textbf{13.88} & 0.10  & 0.03  \\
Font Size + Bold        & 0.540 & 1.81  & 2.00 & 0.66 & 1.61  & 1.13  & \textbf{6.17}  \\
Font Size + Name        & 1.764 & 1.32  & 0.47 & 0.24 & 0.02  & 0.70  & 0.34  \\
Font Name + Bold        & 0.472 & 0.08  & 2.92 & 0.08 & 0.18  & 0.08  & 0.00  \\
Font Size + Name + Bold & 0.244 & 0.21  & 1.26 & 0.03 & 0.03  & 0.09  & 0.13 
\\
\bottomrule
\end{tabular}}
\end{center}
\end{minipage}
\end{table}

The texts were \textit{Tweets}, i.e. messages or status updates on the microblogging site \textit{Twitter}, which are limited to 280 characters, written in German. The tweets were randomly downloaded from the Twitter pages of major news platforms and cleaned up from all mentions and hashtags. In order to ensure that subjects did understand the content of the texts, they were instructed to categorize each tweet after reading. To this end, eight categories were randomly displayed after subjects had completed reading a tweet. Participants were instructed to select as many categories applying to the content of the tweet as they thought appropriate. The tweets had been labelled previously independently by two of the authors of this study. A tweet was considered as labeled correctly by a participant if at least one correct label was selected for each text. A $5 \times 5 \times 2$ factorial design was chosen to investigate the influence of different fonts on reading speed. Each trial consisted of a combination of a different typeface (Arial, Cambria, Century Schoolbook, Helvetica and Times New Roman), a font size (10,20,30,40,50 pt) and whether the bold font was chosen. Each combination was presented exactly five times during the experiment, resulting in a total number of $5 \times 5 \times 2 \times 5 = 250$ trials. The measurement of reading time was started 
by subjects pressing a button, which also revealed a text,  
and stopped by a second button press, which also revealed the screen for the categorization. 
The time returned was used to compute the reading speed as the number of words read per minute. The distribution of all measured reading speeds can be found in Figure \ref{fig:speed-histograms}.

To analyze the influence of the various factors, a Bayesian ANOVA was carried out on a subject by subject basis using JASP \cite{JASP2020}. The Bayes Factors of the models of each subject are shown in Table \ref{tab:preliminary}. While there was substantial evidence in most subjects that font size influenced reading speed, font weight showed a substantial influence only in two subjects. For one subject, the analysis showed substantial evidence for an interaction between font size and bold face.
These results provide evidence for individual differences in the magnitude and directions of the respective influencing factors. The results therefore provide a first indications that the factors influencing reading speed may differ on an individual subject basis.

\subsection{Learning a Font Space}
To generate fonts on a continuum, we first require a parametric generative font model. Here we used unsupervised learning to obtain a continuous font space. 
\begin{figure}[!b]
    \centering
    \includegraphics[width=0.75\columnwidth]{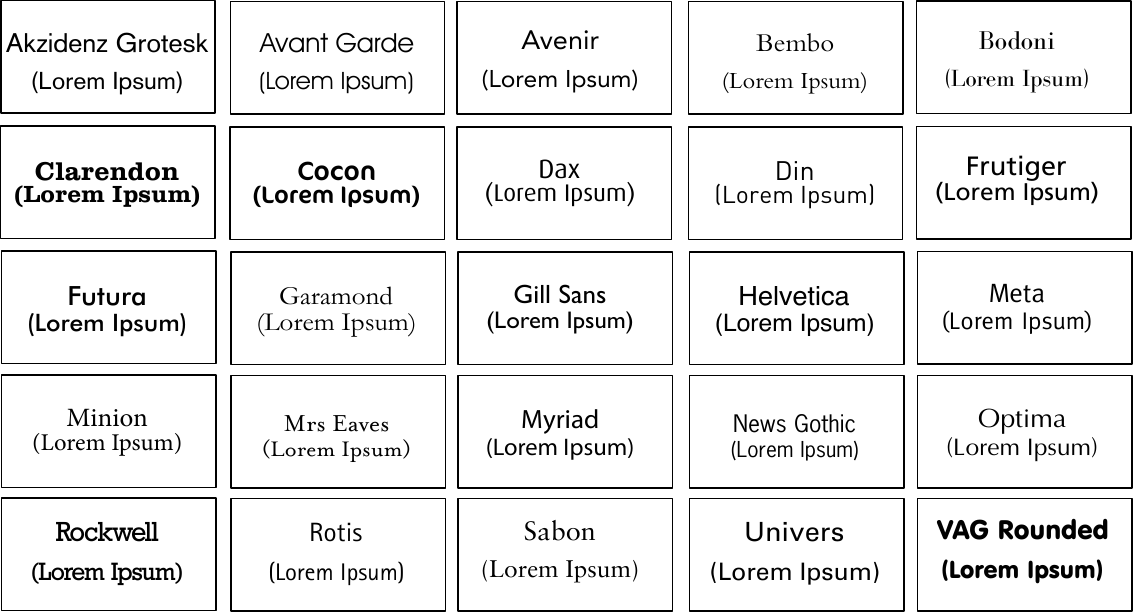}
    \caption{The 25 baseline typefaces used in learning of a font space with NMF.}
    \label{fig:fonts}
\end{figure}

Specifically, we chose non-negative matrix factorization (NMF) \cite{Suvrit2006} as a method for dimensionality reduction of fonts. The idea behind this procedure is to approximate a matrix $\mathbf{X}$ with the product of two matrices $\textbf{W},\textbf{H}$ so that the following relationship applies: $\mathbf {X} \simeq \mathbf {W} \mathbf {H}$. This factorization is done under the constraint, that the approximation minimizes the Frobenius-Norm $||\mathbf{X}-\mathbf{WH}||_{F}$ and that all entries of $\textbf{W},\textbf{H}$ are non-negative. The columns of $\textbf{W}$ then represent the dimensional features and $\textbf{H}$ contains the weights to combine these features to reconstruct the rows in $\textbf{X}$. One advantage of the strictly positive components of NMF with respect to other methods such as Principal Component Analysis or Transformer Networks is 
that the basis functions can be thought of as printer's ink on paper and therefore as elements resembling actual glyphs. 
Another recent approach for font generation employs GANs, see e.g. \cite{GAN1,GAN2, GAN3, GAN4, GAN5}. Even though this technique gives good results at the letter and word level,
current implementations suffer from 
problems in alignment and kerning in continuous texts. Due to these difficulties, 
we were not able to employ
GANs that could generate qualitatively satisfactory texts even though this approach will certainly be very promising in the future.

\begin{figure}[!b]
\centering
\begin{minipage}{.48\textwidth}
  \centering
  \includegraphics[width=0.98\textwidth]{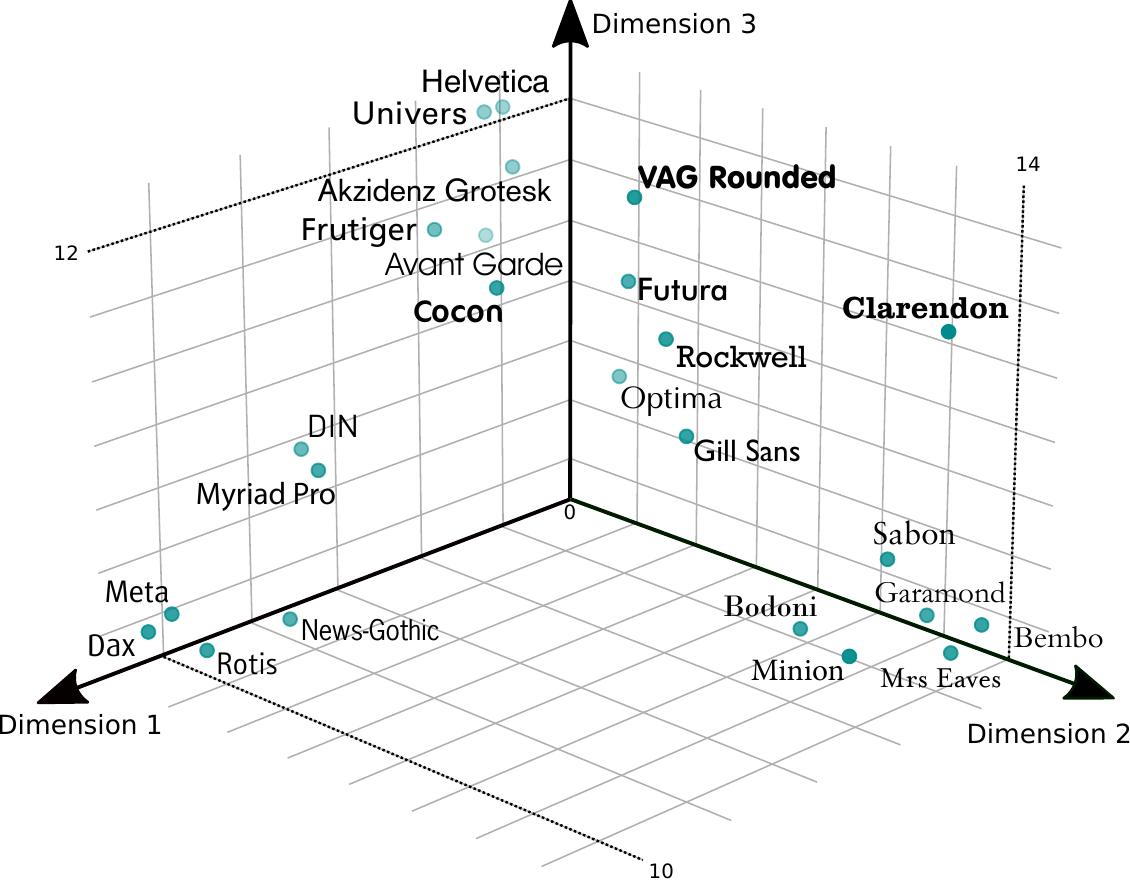}
  \captionof{figure}{Three-dimensional representation of the original fonts in the learned NMF font space.}
    \label{fig:fontspace}
\end{minipage} \hfill
\begin{minipage}{.49\textwidth}
  \centering
  \includegraphics[width=0.98\textwidth]{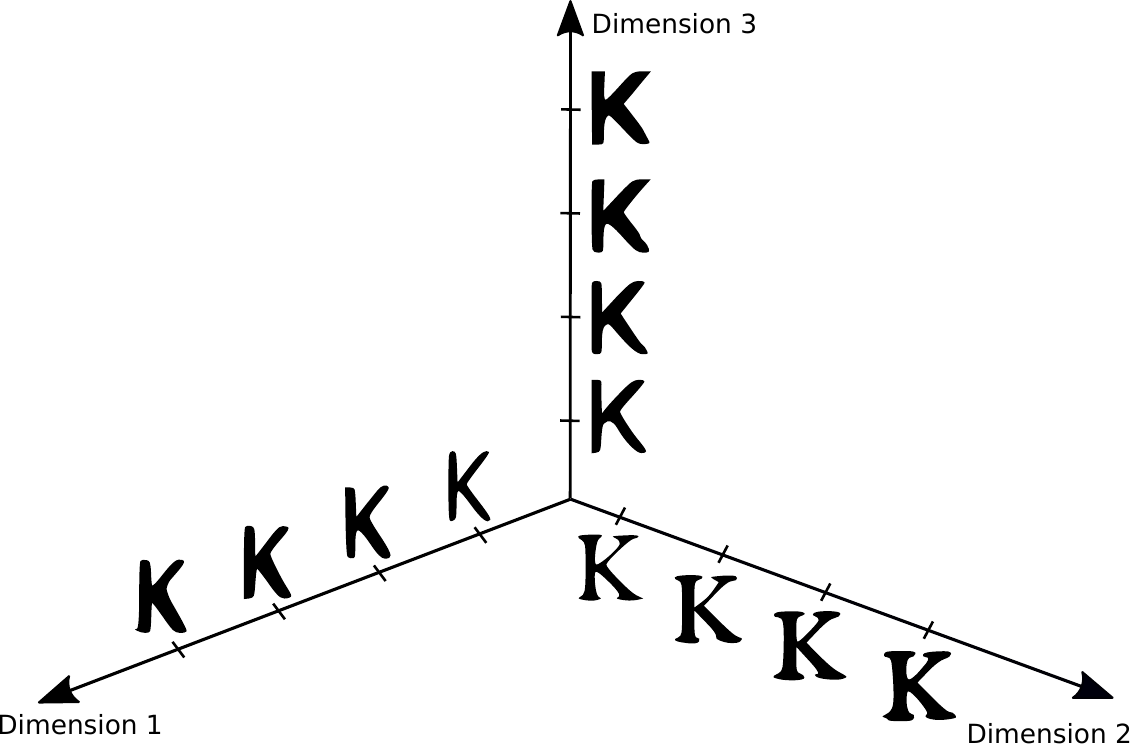}
  \captionof{figure}{Influence of the three font dimensions exemplarily demonstrated for one capital letter K.}
    \label{fig:dimensions}
\end{minipage}
\end{figure}

In order to cover a basic set of fonts including fonts with serifs and sans-serif fonts, we selected a list of classic and popular typefaces \cite{ClassicalFonts}, containing a total of 25 \textit{classical} fonts. The typeface names are shown in their respective font in Figure \ref{fig:fonts}.
To perform the NMF, we generated a grayscale image of size $2375 \times 51$ pixels containing
all 26 letters in upper and lower case, the numbers from zero to nine, the German umlauts, brackets, question and exclamation mark, dot, comma, hyphen, colon, semicolon, slash and quotation marks. 
Such an image was generated for each individual font using a font-size of 40 pt. Letters were arranged side by side.
The image data was then concatenated together with information about the alignment of each glyph in the font, obtained from the respective information in the TrueType-file. The resulting font vectors form the rows of our matrix \textbf{X}, on which we performed the NMF.

To choose the appropriate number of dimensions, cross-validation was performed. As recommended by Kanagal \& Sindhwani \cite{WeightedNMF}, we masked our data using Wold holdouts and fitted the NMF to the rest of the data using weighted NMF with a binary weight matrix. The Wold holdouts are created by splitting the matrix into 16 blocks and randomly selecting four blocks, one for each row, to hold out. Ten random Wold holdouts where created and for each we calculated the reconstruction error when choosing between one and five NMF components.

The mean reconstruction error showed that a low testing error can be obtained between one and three NMF components and that the model starts overfitting the data for four or more components. For our subsequent experiments we therefore decided to use three components to ensure a sufficiently rich font representation, while also avoiding overfitting to the data. The restriction to three dimensions had the additional advantage of allowing an individual font to be represented by a very low dimensional vector with only three entries, which reduces the computational burden for the subsequent Bayesian optimization process described in the following. The representation of the 25 fonts in the three-dimensional NMF space is shown in Figure \ref{fig:fontspace}.

\begin{figure}[!t]
    \centering
    \includegraphics[width=0.75\columnwidth]{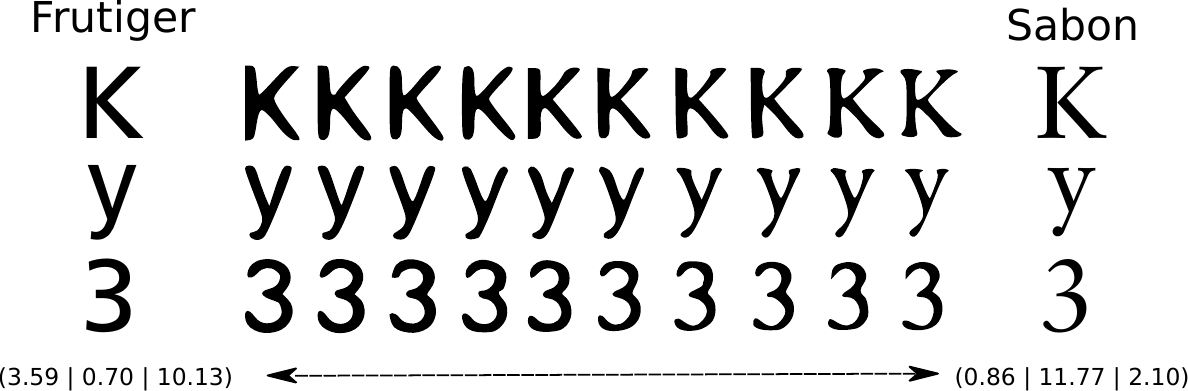}
    \caption{Interpolating between the two fonts Frutiger and Sabon. At the edges are the original fonts, right next to them the NMF approximations and in between the linear interpolation through the font space. The coordinates for both fonts in our three-dimensional space is given.}
    \label{fig:tdg}
\end{figure}

Through this representation it is now possible to synthesize new fonts as a linear combinations of the three learnt font basis vectors. Accordingly, a font can be represented by a point in this three dimensional space. Figure \ref{fig:dimensions} shows exemplary the influence of the three dimensions for the upper case letter K. Inspecting the letter renderings suggest that
the first and third dimension are related to the scaling of letters in the vertical or horizontal directions, whereas the second dimension is related to the presence and strength of serifs.
This generative model allows not only creating new fonts, but also the interpolation between actual fonts. Figure \ref{fig:tdg} shows the changes in three characters (capital \textit{K}, lower case \textit{y} and number \textit{3}) resulting from linearly moving in euclidean space between the points corresponding to the fonts Frutiger and Sabon. Note that the changes are gradual and smooth. 

To be able to generate texts with a synthesized font we generate TrueType-Font (TTF) files on the fly. A linear combination of the  basis vectors obtained through NMF yields a vector that contains a new greyscale bitmap image of all the glyphs in the new synthesized font as well as alignment information for each individual glyph. The bitmap images of the glyphs are traced individually to Scalable Vector Graphics (SVG) using Potrace \cite{Potrace} and then stitched together into an SVG font. The SVG font then was converted into the TTF format using FontForge \cite{FontForge}. At this point, the alignment information is directly inserted into a TTF file using the fontTools python package \cite{FontTools}. The entire process of font generation works automatically and in real time. 

\subsection{Bayesian Optimization}
To generate new fonts that increase participants' reading speed, we need to select an optimization technique that finds corresponding regions in the three dimensional font space. The underlying assumption is that fonts change smoothly within the generative font space and that individual's reading speed is also changing smoothly along similar fonts.
The objective function to be maximized is an individual's reading speed as function of a specific font, which in this case is represented by a point in our three dimensional font space. Since the font space is infinitely large and only a limited number of texts can be represented, we decided to use Bayesian Optimization, because it is a well-known optimization technique to find extrema of complex cost functions when the number of samples that can be drawn is limited \cite{brochu2010tutorial}. 
Bayesian optimization starts with an a priori uncertainty across the three-dimensional font volume and selects successive points within that volume, for which the reading speed is evaluated through a reading experiment. Since the choice for this prior distribution is hard in general, a common choice is the Gaussian process prior due to its flexibility and tractability \cite{Snoek2012}. The central assumption of the Gaussian Process is that any finite number of points can be expressed by an Multivariate Gaussian Distribution. The idea now is to take one of these points and assume that it is the value of the underlying unknown function. With the marginalization properties of the Multivariate Gaussian distribution it is now possible to compute marginal and conditional distributions given the observed data point. By obtaining the reading speed for this font through an experiment  the algorithm reduces the uncertainty at that location in font space. To select the next best font for testing, a balance has to be struck between exploration, i.e. selecting a region in font space for which the uncertainty about reading speed is large, and exploitation, i.e. selecting a region in font space where the reading speed is expected to be high, given previous reading experiments. This selection process is handled by the so called acquisition function and its corresponding parameters. Thus, Bayesian Optimization proceeds by sampling more densely those region, where reading speed is high, and more evenly in those regions, where uncertainty is high.

\subsection{Experimental Design and Data}
The overall closed-loop logic of the experiment (see Figure \ref{fig:teaser}) was to have the Bayesian optimization algorithm generate a font in the generative font space and use the reading speed of individual participants to generate new fonts to increase the reading speed.  

\subsubsection{Participants}
Eleven subjects (5 females, 6 males; age $M$ = 24, $SD$ = 2.64) participated in the experiment. All participants were German native speaker and had normal or corrected to normal vision. Due to the COVID-19 pandemic, it was not possible to invite subjects to the laboratory, so they all performed the experiment on their private computers at home. Participants were acquired from graduate and undergraduate students of the research group and received the necessary materials for participation by e-mail. All subjects received a detailed, multi-page instruction to keep influences like sitting position, viewing distances, room lighting, etc. as similar as possible. In addition to detailed instructions and information on the experiment, the subjects also received an executable file containing the experiment. Subjects were naive with respect to the mechanics of font generation and selection.

\subsubsection{Stimuli}
Subjects read a total of 95 texts, which were taken from a German Wiki for children's encyclopedia texts. These texts were chosen because they 
are
easily understandable for adult native speakers, so that 
the content of the texts had minimal effect on reading speed.
Furthermore, the texts were chosen so that they were comparable in length, i.e. the number of words for the first 94 texts was on average 99.7 with a minimum of 91 and a maximum of 109 words. Additionally, a single text with only 51 words was selected to check whether the reading speed deviated significantly depending on text length. The texts were presented in a random order for each participant.

\subsubsection{Procedure}
Subjects were instructed to read the texts attentively and as quickly as possible but only once in total. To check that the texts were read and processed in terms of content, the subjects had the task of detecting words from previously specified categories while reading. Each individual trial started with the introduction of the category for the next text, e.g. before reading the text, it was indicated that words from the category \textit{animals} should be detected subsequently. The category only referred to the next text, and a new category was selected for each trial. 
Each text's category had been previously independently labelled by two of the authors of this study and each subject was given the same category for a text.

\begin{figure}[!b]
    \centering
    \includegraphics[width=0.65\columnwidth]{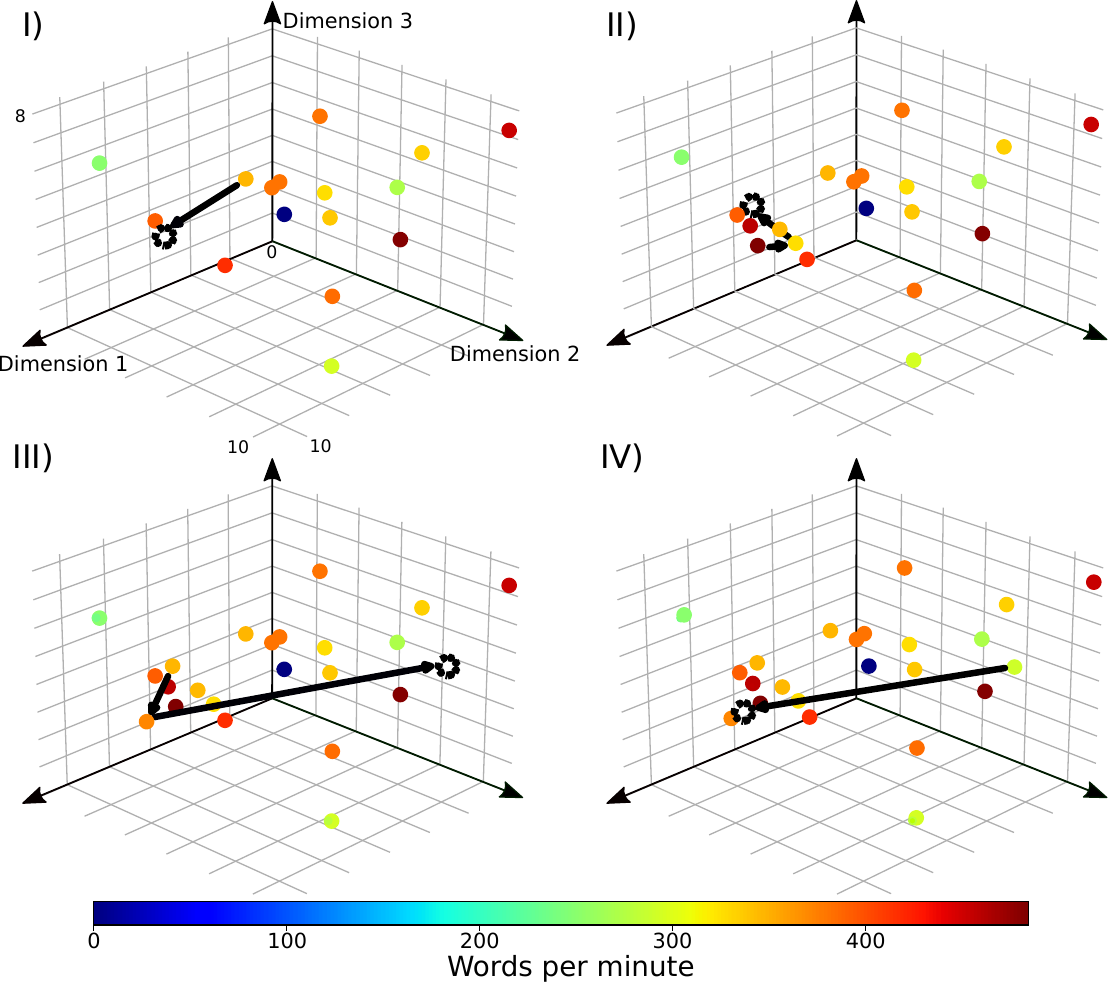}
    \caption{Illustration of the Bayesian optimization probing process, after a new maximum was found. In I) and II) new points in a high reading speed cluster are sampled in an exploitation phase. After a slow new sample in III) was found in the exploration phase, the process is searching again in the region of the faster speeds in IV). Data from subject 7 in trials (I) 19 (II) 23 \& 24 (III) 25 \& 26 (IV) 27 respectively.}
    \label{fig:bayes_move}
\end{figure}

For each text there 
were between one and ten words belonging to the instructed categories (mean 3.07, SD 2.05). Once a subject had read the category for the next trial, they could use the space bar to display the text and begin reading. When the text was displayed, the time measurement for the reading speed also started. To avoid an accidental start, the task could only be initiated a few seconds after the category had been displayed (this was indicated to the participants by a red/green signal). Each time a term matching the previously specified category was detected, the participants could press the space bar to indicate this. As an example, in the short sentence "The quick brown fox jumps over the lazy dog.", when reading the words \textit{fox} and \textit{dog}, the user had to press the space bar to indicate that the objects were recognized for the category \textit{animals}. Once they had finished reading the text they could press the Enter key to stop reading and thus also stop the timing.

In addition, a multiple choice question was asked after having read eleven texts at random trials throughout the experiment, in order to additionally test subjects' text comprehension. The question always referred to the last text read and six possible answers were given, of which exactly one was correct. After solving one of the multiple choice questions, participants received feedback on their answer. This feedback consisted of the average reading speed, the number of correct detections, and the correctness of the multiple choice question. These three components were combined into a score to further motivate participants to read quickly, but also correctly and attentively.

In order to investigate whether regions of the generative font space exist which allow higher reading speeds, subjects read the 95 texts in different synthesized fonts. Since the three-dimensional space is infinitely large and only a limited number of texts can be presented, the Bayesian global optimization with Gaussian process was used in order to sample only 95 fonts. The target function to be optimized is the reading speed as parameterized by the three dimensions. The process was implemented with \cite{BayesOpt} and the following configuration was selected: We choose the Upper Confidence Bound (UCB) as acquisition function for balancing exploration and exploitation of new fonts \cite{UCB}. The exploration parameter was set to $\kappa=5$. We started with ten random initialization points and chose the noise handling parameter as $\alpha=0.001$ for the exploration strategy. We chose a Matern Kernel with parameter $\nu=2.5$ for the covariance function and the L-BGFS-B algorithm for the optimization of the kernel parameters. Figure \ref{fig:bayes_move} shows an example for the sampling process through the Bayesian optimization algorithm.

As combinations of NMF components that have too high or too low magnitudes would for sure lead to unreadable fonts, we constrained the exploration space. If the sum of the three components is too low, the font is only faintly visible. If instead it is too high, the font is too bold and not readable either. Therefore, we empirically decided to constrain the magnitude of every dimension to be between 0 and 13 and the sum of all magnitudes to be between 7 and 20 units.
It could happen that, in spite of these constraints, generated fonts were not readable. In such cases, subjects had been instructed to reset the corresponding trial via a previously defined key. Afterwards, the same text was presented in a new font, ensuring that all participants had read all 95 texts. This happened on 94 trials in total across all subjects (mean 8.54, SD 3.79).
The Bayesian optimization algorithm nevertheless received the data that this linear combination returned a reading speed of zero words per minute.

\section{Results}

\subsection{Ruling out Confounders}
\label{confounders}

In order to rule out the possibility that the texts, despite the selected constraints, led to differences in reading speed in terms of content, the average reading speed in words per minute was determined for each text over all participants. A Bayesian ANOVA was used to check whether there was a significant difference in mean value and thus a dependence of the reading speed on the respective text. The Bayes Factor $B_{01}=360.36$ gives decisive evidence for the null hypothesis that there was no significant differences between the texts. 

A second possible confounder relates to the length of the texts, which participants read. By including a text with 51 words, we were able to check that the reading speed as measured in words per minute did not significantly depend on the chosen length of texts. Indeed, the reading speed for the shorter text (mean:  257.37 words per minute) was within the 95\% confidence interval of reading speeds of the 94 remaining texts for all individual participants (mean: 265.02, SD: 89.82).

Another confounder refers to the detection task and the number of key presses required during a trial. In order to investigate whether the detections and motor behavior have an influence on the measured time, the Bayesian Pearson Correlation Coefficient between the number of words belonging to the instructed category and the corresponding reading time was computed. For this purpose the reading speeds were averaged over the number of expected detections to exclude the influences of all other factors influencing reading speed. No significant correlation was found ($r=-0.223;BF_{10} = 0.47$) thereby confirming that the number of words belonging to the instructed category did not have a significant effect on reading speed.

\begin{figure}[!b]
    \centering
    \includegraphics[width=0.5\columnwidth]{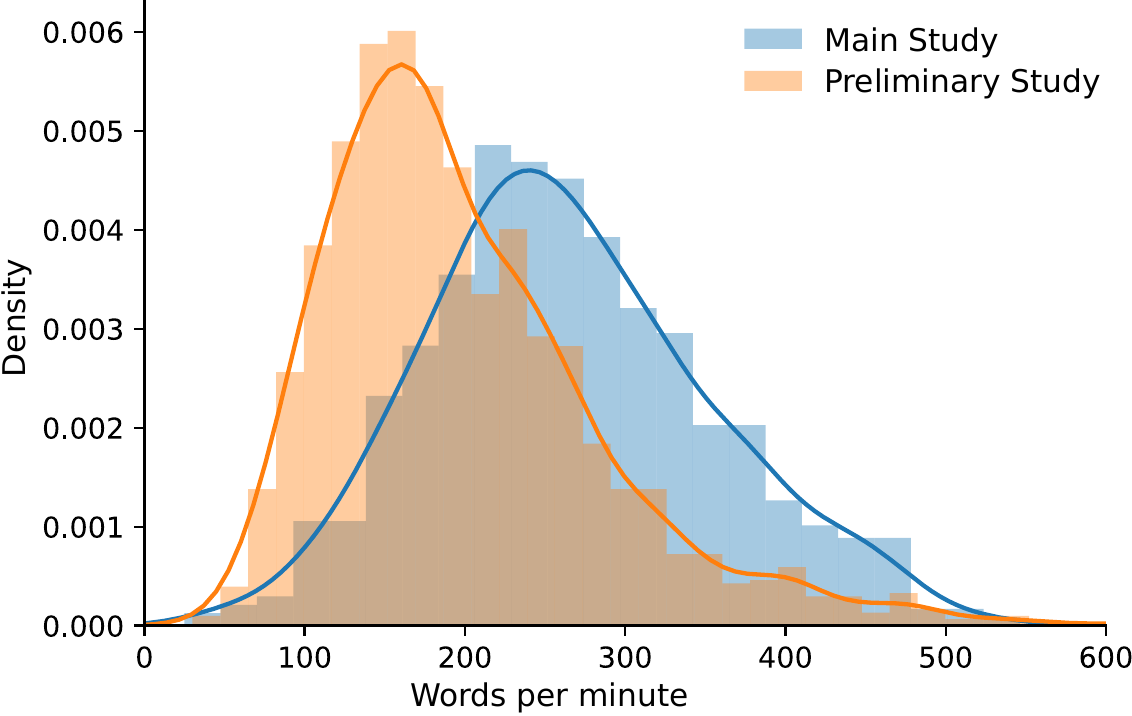}
    \caption{Histograms of measured reading speeds for the preliminary (orange) and the main study (blue). A kernel density estimator was fitted to both histograms.}
    \label{fig:speed-histograms}
\end{figure}

It might be argued that although reading speeds with AdaptiFont may be improved, reading speeds are still larger for traditional fonts. To exclude this possibility, we compared the reading speeds over all texts and subjects between the preliminary study, in which classic typefaces were used, and the main study using AdaptiFont. A Kolmogorov-Smirnov test indicates that the two distributions are indeed significantly different ($D=0.367;p<.001$) and a Bayesian Two-Samples t-Test showed that the mean reading speeds for AdaptiFont were significantly higher than for the traditional typefaces ($BF_{+0}=6.782$).
Figure \ref{fig:speed-histograms} shows the corresponding distributions. 

Finally, we asked whether the optimization throughout the experiment actually reduced uncertainty about reading speed within the font space. For this, the point to be sampled next by the optimization process was considered. At this point, the variance of the Gaussian Process was integrated within the neighborhood of a font with radius $r=0.225$ before and after reading a text in the corresponding font. This radius is exactly half the distance between the two closest classic baseline fonts. For each iteration, the variance change in this region was now examined. Bayesian optimization indeed resulted in an average variance reduction of 943.65 units per iteration, the corresponding distribution can be found in Figure \ref{fig:uncertainty}.

\subsection{Clustering}

To detect regions in the generative font space with higher reading speeds a clustering method was used, which includes the reading speed as fourth dimension in addition to the three font dimensions for each individual participant. We chose the OPTICS algorithm \cite{OPTICS} to cluster the data points, which is a density based clustering procedure, which orders the points of a data set linearly, so spatially nearest neighbours can be clustered together. This has the advantage over other methods, such as k-Means, because the number of clusters does not have to be specified in advance and allows detecting outliers that are not assigned to any cluster. 
The latter feature was useful in the analysis of our data, as it excluded trials with deviations in reading speed due to lapses in attention.

\begin{figure}[!b]
    \centering
    \includegraphics[width=0.65\columnwidth]{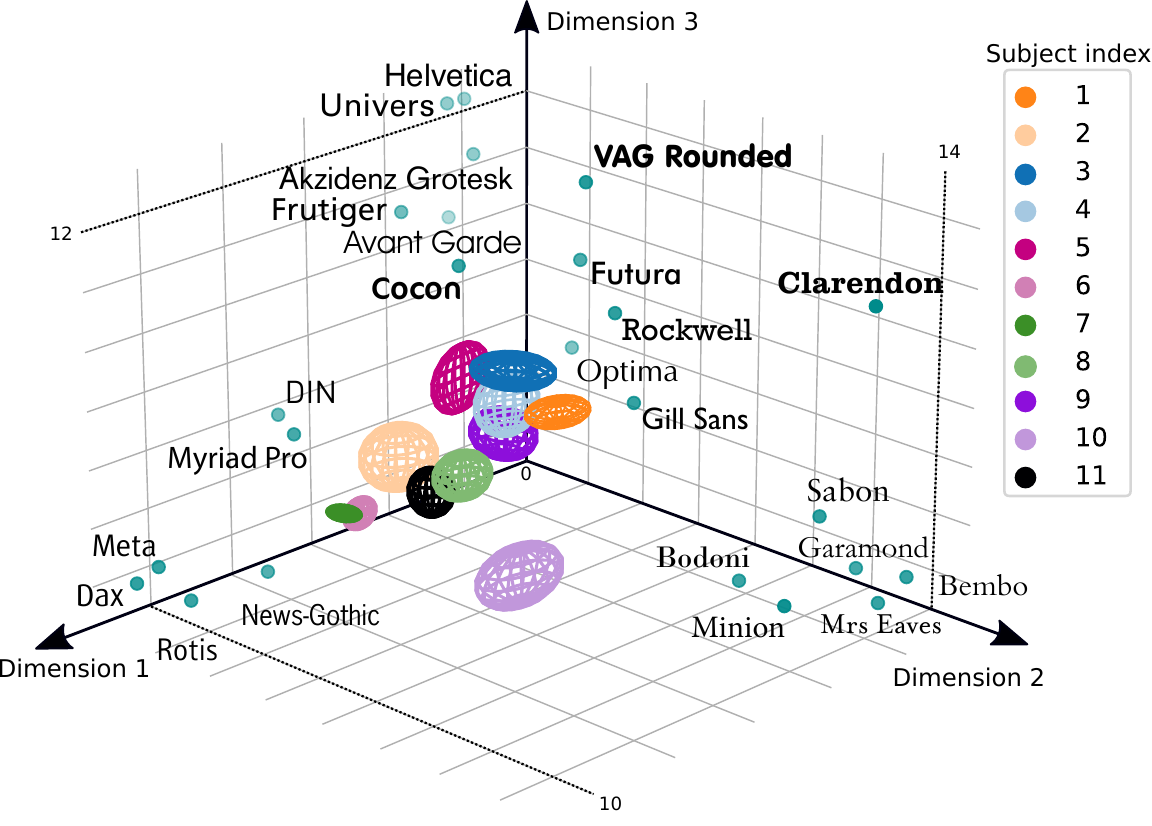}
    \caption{Ellipsoids with centroids and standard errors in the three font space dimensions of clusters corresponding to highest reading speed.}
    \label{fig:cluster_space}
\end{figure}

As a free parameter of the algorithm, the minimum number of data points that must be present in a cluster must be defined. Following the recommendations of \cite{GDBSCAN}, this parameter can be determined using the heuristic of using twice the number of data dimensions minus one, i.e. $(2 \times \text{Dimensions})-1$, so that in our case we decided to set the parameter to $n=5$. For visualization purposes, Figure \ref{fig:cluster_space} shows the best clusters, i.e. the clusters with the highest mean average reading speed for each of the 11 subjects in the generative font space. The clusters are represented by an ellipsoid, with the center of all associated data points and the standard error in all three dimensions as main axes.

In order to better compare the optimized synthesized fonts with the traditional fonts, we additionally considered their position in font space with respect to each other. The optimized fonts for all subjects lay within a region bounded by the traditional fonts Rockwell, Myriad, Optima and News Gothic. To obtain an indication of the similarity and difference between individuals' optimized font, we computed the mean pairwise distance between individuals' optimized fonts. This distance is 3.99 units in the font space, which is comparable to the distance between Avantgarde and Universe.
By comparison, the distance between slowest and fastest fonts for each subject was on average 11.3 units. 
In order to compare these distances within the font space, we computed the distances between classic fonts, with the smallest distance (0.45) between Helvetica and Univers, the largest distance (18.8) between Univers and Bembo, and the average distance (9.96), which was comparable to the distance between Optima and Minion.

\subsection{Reading Speed Improvements with AdaptiFont}
\begin{figure}[!t]
    \centering
    \includegraphics[width=0.65\columnwidth]{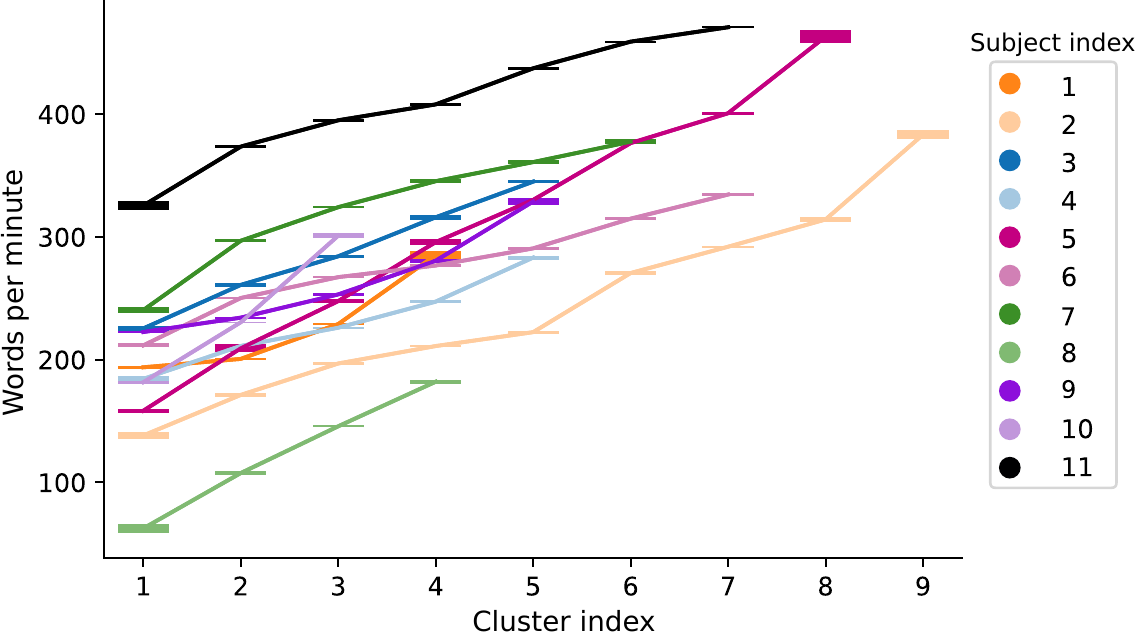}
    \caption{Found clusters for every subject based on the OPTICS algorithms and the corresponding mean reading speed together with their standard errors. The mean reading speeds of all cluster were rank ordered for each subject. Note that the standard errors of reading speed were very small compared to the mean reading speeds for each cluster and subject.}
    \label{fig:clusters}
\end{figure}

To check whether the font clusters correspond to significantly different reading speeds, a Bayesian ANOVA was calculated. Figure \ref{fig:clusters} shows the mean reading speeds for  each of the found font clusters together with the corresponding 95\% credible interval for each subject.
The Bayes Factors are reported in Table \ref{tab:modelComparison}, where every Bayes Factor gives decisive evidence against the null hypothesis, so that the clusters differ significantly in their average reading speed. 
A MANOVA established statistically significant differences in reading speed between 
clusters of highest reading speed of the subjects, $F(10,75) = 3.24$, $p < .001$; $\text{Pillai Trace} = 0.91$. The corresponding ANOVA for every single dimension was performed resulting in three significant differences $D_1:$ $F(10,75)=2.01, p<.05$; $D_2:$ $F(10,75)=3.63, p<.05$; $D_3:$ $F(10,75)=4.02, p<.05$. 
Thus, the differences between clusters of individuals' highest reading speed visible in figure \ref{fig:cluster_space} are statistically significant.
To summarize together with the results from section \ref{confounders} and Figure \ref{fig:speed-histograms}: The distribution generated with AdaptiFont is not only statistically significantly different, the reading speed with Adaptifont is also statistically significantly faster on average. All subjects showed improved reading speed.

\begin{figure}[ht]
    \centering
    \includegraphics[width=0.65\columnwidth]{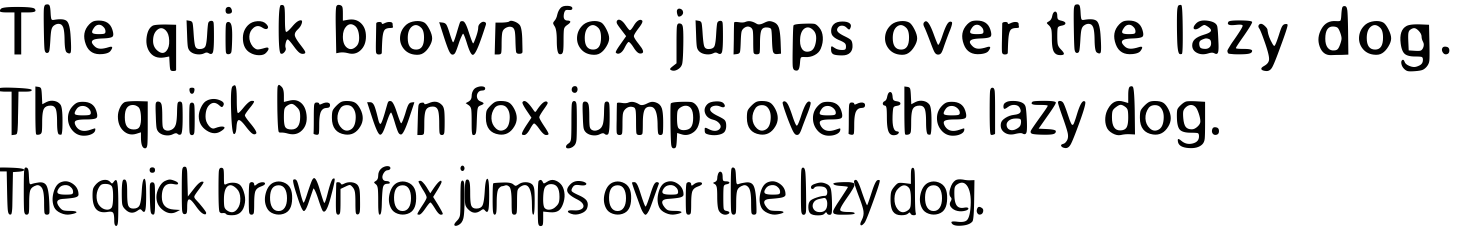}
    \caption{Fonts generated from the centroids of the best clusters of subjects 1, 5, and 6.}
    \label{fig:sub_fonts}
\end{figure}

Finally, figure \ref{fig:sub_fonts} shows a pangram written in the respective font of the center of the cluster of highest reading speed for three exemplary subjects. The fonts are actual TrueType-fonts and can be used as such.

\section{Discussion and Limitations}

We presented AdaptiFont, a human-in-the-loop system to increase readability of digitally rendered text. The system uses a learned generative font space and Bayesian optimization to generate successive fonts that are evaluated in terms of individual user's reading speed.
A preliminary study gave first indications that 
individual differences in readability exist in terms different font features.
With AdaptiFont, variation in font features are captured in a specific font space learned through NMF. 
Using this representation, we explore the reading speed of individual users with the help of Bayesian optimization.
The results of the user study show the feasibility of the approach, both in terms of the required data to find fonts that increase individual's reading speed as well as 
the statistically significant magnitude of the improvement in individual's reading speed. Finally, significant differences between subjects exist in the font regions that contain fonts that are associated with high legibility.
The generated fonts are actual TrueType-font files, which can be installed and used.

Although reading speed maximizing fonts found in this study were significantly different across individual subjects, it cannot be concluded that our system finds one single font that maximizes reading speed unchangeably for a single subject or across all subjects. Rather, our system can be understood as dynamically and continuously creating fonts for an individual, which maximizes the reading speed at the time of use. This may depend on the content of the text, whether you are exhausted, or perhaps are using different display devices. The empirical data we obtained in the experiments provides clear evidence that Adaptifont increased all participants’ reading speeds. How these optimized fonts are related across subjects or within subjects across days or viewing contexts are very interesting further empirical questions, which can be systematically addressed by utilizing Adaptifont in future research.

While the current study demonstrates the feasibility of interactively optimizing fonts through Bayesian optimization, it has a number of limitations, which should be addressed in further research. 
First, while reading speed is a natural target for optimizing fonts, other criteria are conceivable, such as text comprehension, aesthetic appeal, or memorability either of texts or fonts.
Second, ample research has demonstrated cognitive and linguistic influences on reading speed, which were not taken into account here. Third, future generative font spaces, e.g. based on GANs, may provide a more expressive font synthesis with smaller approximation errors to classic fonts and better alignment and kerning. Fourth, while the current study maximized reading speed at the individual subject level, it is straightforward to use the system to maximize optimization criteria across multiple subjects. Fifth, the generative font space based on NMF may result in synthetic fonts that violate typographic rules. Including typographic constraints or interactively including a typography expert in the font synthesis process may yield an additional avenue for the development of new fonts.

\section{Conclusion}
While constant, variable, and parametric fonts have been designed and developed in the past, Adaptifont is an interactive system, which generates new fonts based on a user's interaction with text. In this paper, we used a generative font model and Bayesian optimization to generate fonts, which
progressively increased users' reading speeds.
Adaptifont can be understood as a system that dynamically and continuously creates fonts for an individual, which in this study maximized the reading speed at the time of use.
Evaluation of the system in a user study showed the feasibility of the approach.
Adaptifonts can be utilized to address further scientific questions such as how the optimized fonts depend on sensory, linguistic, or contextual factors or how optimized fonts are related across individuals. 
The system does in no way detract from other work on generating fonts,
both from the point of view of developing such algorithms or 
from the point of view of the typographer,
as developing typefaces can be guided by different motivations and with different goals.

\section*{Code}
Under \url{https://github.com/RothkopfLab/AdaptiFont} you can find the repository with additional material. Provided are usable fonts in the TrueType-font (.ttf) format from the centroids of the clusters with highest reading speed for all subjects, as well as a script that can be used to generate fonts from the learned NMF components. 

\section*{Acknowledgements}
We would like to thank Nils Neupärtl and David Hoppe for their constructive comments. The work is funded by German Research Foundation (DFG, grant: RO 4337/3-1; 409170253).

\bibliographystyle{unsrt}
\bibliography{references}

\appendix
\renewcommand\thefigure{\thesection.\arabic{figure}}    
\section{Appendix}
\setcounter{figure}{0} 

\begin{figure}[ht]
    \centering
    \includegraphics[width=0.5\columnwidth]{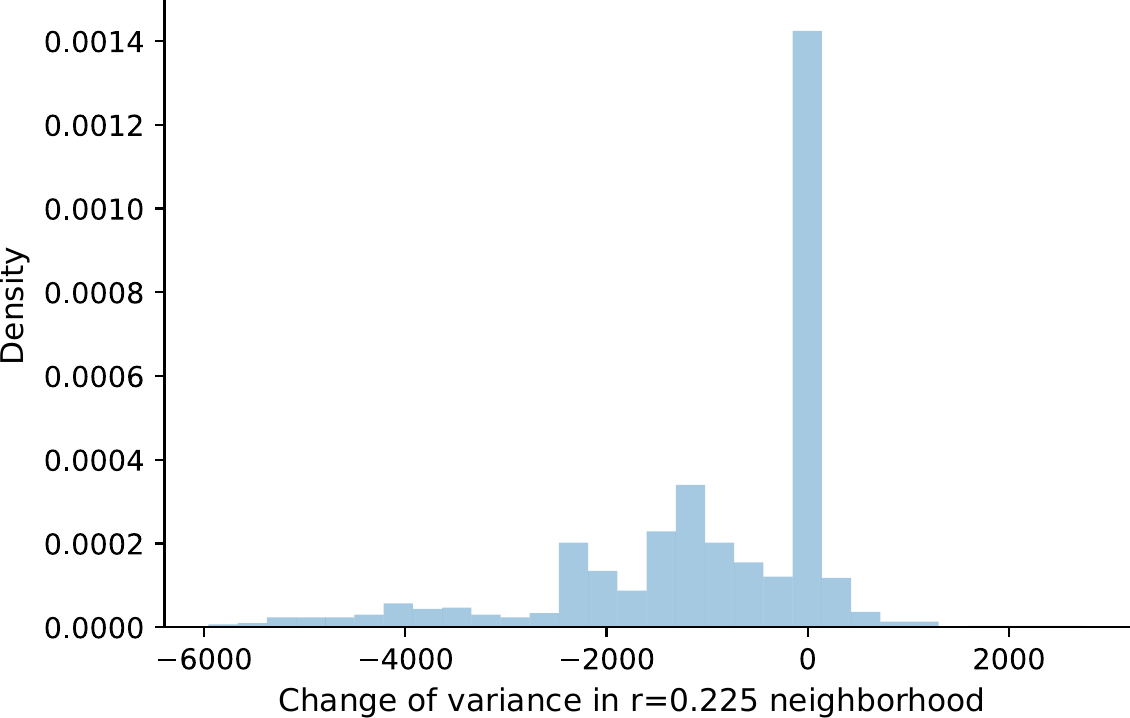}
    \caption{Distribution of the per-sample uncertainty reduction. The histogram shows the distribution of all changes in uncertainty about the individual user's reading speed in the neighborhood of a sampled font after a reading experiment with a single font. }
    \label{fig:uncertainty}
\end{figure}

\begin{table}[ht]
	\centering
	\caption{Bayes Factors of the ANOVAs relating individuals' reading speed between the found clusters.}
	\label{tab:modelComparison}
	{
		\begin{tabular}{lr} 

			Subject &  BF$_{10}$   \\
			\cmidrule[0.4pt]{1-2}
			1 & 4.458e+23  \\
			2 & 5.651e+56  \\
			3 &  5.890e+35  \\
			4 & 3.592e+30   \\
			5 & 5.752e+45  \\
			6 & 6.091e+41  \\
			7 & 1.804e+37  \\
			8 & 5.974e+31   \\
			9 & 4.683e+29  \\
			10 & 5.956e+14   \\
            11 & 2.008e+48  \\
			\bottomrule
		\end{tabular}
	}
\end{table}

\end{document}